\documentclass[twocolumn,pre,class]{revtex4-1}%
\usepackage{amsmath}
\usepackage{graphicx}
\usepackage{bm}
\usepackage[colorlinks,linkcolor=blue,anchorcolor=blue,citecolor=blue,urlcolor=blue,dvipdfm]{hyperref}
\begin{document}

\title{Detection of Single Nanoparticles Using the Dissipative Interaction in a High-$Q$ Microcavity}
\author{Bo-Qiang Shen$^{1}$}
\author{Xiao-Chong Yu$^{1,2}$}
\author{Yanyan Zhi$^{1,2}$}
\author{Li Wang$^{1,2}$}
\author{Donghyun Kim$^{3}$}
\author{Qihuang Gong$^{1,2}$}
\author{Yun-Feng Xiao$^{1,2}$}
\email{yfxiao@pku.edu.cn}
\homepage{http://www.phy.pku.edu.cn/~yfxiao/}

\affiliation{$^{1}$ State Key Laboratory for Mesoscopic Physics and School of Physics, Peking University, Beijing 100871, People¡¯s Republic of China}
\affiliation{$^{2}$ Collaborative Innovation Center of Quantum Matter, Beijing 100871, People¡¯s Republic of China}
\affiliation{$^{3}$ School of Electrical and Electronic Engineering, Yonsei University, Seoul, 120--749, South Korea}

\date{\today}

\begin{abstract}
Ultrasensitive optical detection of nanometer-scaled particles is highly desirable for applications in early-stage diagnosis of human diseases, environmental monitoring, and homeland security, but remains extremely difficult due to ultralow polarizabilities of small-sized, low-index particles. Optical whispering-gallery-mode microcavities, which can enhance significantly the light-matter interaction, have emerged as promising platforms for label-free detection of nanoscale objects. Different from the conventional whispering-gallery-mode sensing relying on the reactive (i.e., dispersive) interaction, here we propose and demonstrate to detect single lossy nanoparticles using the dissipative interaction in a high-$Q$ toroidal microcavity. In the experiment, detection of single gold nanorods in an aqueous environment is realized by monitoring simultaneously the linewidth change and shift of the cavity mode. The experimental result falls within the theoretical prediction. Remarkably, the reactive and dissipative sensing methods are evaluated by setting the probe wavelength on and off the surface plasmon resonance to tune the absorption of nanorods, which demonstrates clearly the great potential of the dissipative sensing method to detect lossy nanoparticles. Future applications could also combine the dissipative and reactive sensing methods, which may provide better characterizations of nanoparticles.
\end{abstract}

\maketitle

\section{Introduction}

Single nanoparticle detection is of critical importance in applications of human disease prediagnosis, real-time environmental monitoring, and semiconductor manufacture control. For example, the detection of a fatal virus with a size ranging from tens of nanometers to a few micrometers \cite{59,60} is a prerequisite for early-stage disease diagnosis, such as AIDS, SARS, and Ebola. Particles with a size less than 300 nm in air can penetrate the lung and sequentially enter the blood, causing severe organ damage. Optical microcavities featuring high-$Q$ factors and small mode volumes, such as Fabry-Perot cavities \cite{62}, photonic crystals \cite{PhC1, PhC2}, microspheres \cite{1,2,3,4}, microrings \cite{5,6,7,8,9}, microtoroids \cite{10,11,12}, microbubbles \cite{13,14,15}, and microtubes \cite{55,56} have been widely investigated in sensing applications.  In general, the microcavity sensing depends mainly on reactive (i.e., dispersive) interactions, resulting in a resonance wavelength shift \cite{16,17,53,18} or mode splitting \cite{18,19}, which essentially responds to the real part of the polarizability of the targets. Via reactive sensing, a single virus \cite{20,21} and a single nanoparticle \cite{22,23,24,25} have been detected experimentally. Furthermore, by employing surface plasmon resonance (SPR) enhancement \cite{26,27,28,29,30,31,32,33}, microcavity lasing \cite{34,35,36,37}, the noise suppression technique \cite{38}, or exceptional point \cite{39}, a better sensing ability can still be achieved.

In many applications, target analytes in the surrounding medium have small polarizabilities which, unfortunately, produce the weak reactive interaction with cavity modes, but may strongly absorb the probe light and then change the linewidth of the cavity mode significantly \cite{10,32,40,41,42,43,44}, representing an effective dissipative sensing method for single nanoparticle detection. Furthermore, combining the dissipative and reactive sensing methods, more information of the analyte can be obtained, and those two kinds of signal may verify with each other and confirm the reliability of the detection results. Note that the linewidth change induced by the absorption loss differs from the mode broadening resulting from additional scattering loss and unresolvable splitting in the optical spectrum \cite{18,21,45}. In this work, we thus propose and demonstrate to detect single lossy nanoparticles via the dissipative interaction with a toroidal optical microcavity which supports high-$Q$ optical whispering-gallery modes (WGMs). Experimentally, gold nanorods acting as the lossy analytes are used to examine the performance of the present dissipative sensing method. Thus, by setting the probe laser wavelength on and off the surface plasmon resonance of the gold nanorods, respectively, both the linewidth change and the shift of cavity mode can be monitored simultaneously in real time to evaluate the reactive and dissipative sensing methods, highlighting the benefits of the later scheme to detect single lossy particles.

\section{Experimental Method}

\begin{figure*}
\includegraphics[width = 14cm]{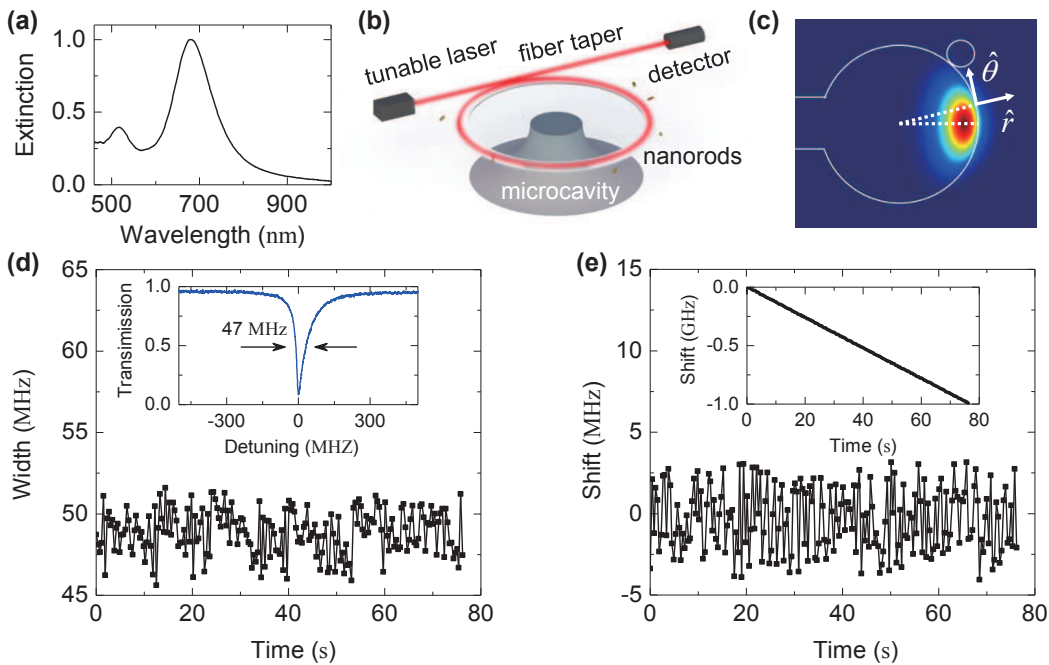}
\caption{(a) Normalized extinction of gold nanorods dispersed in water, showing the plasmonic resonance at 680-nm wavelength. (b) Schematic drawing of the experimental setup. A tunable laser is coupled into a high-$Q$ microtoroid by a fiber taper. (c) The normalized cross-sectional distribution of a transverse electric (TE) WGM, and the relative position of the taper and the cavity.(d) The measured linewidth of a WGM in deionized water, showing an uncertainty of $1.35$ MHz. The inset plots the transmission spectrum of the WGM, giving an intrinsic $Q$ factor of $2\times10^7$. (e) The measured mode shift of the same WGM, with an uncertainty of $1.96$ MHz after subtracting a large background frequency drift (inset).}\label{fig1}
\end{figure*}

\emph{Fabrication of microtoroid cavity.} Silica microtoroids are fabricated sequentially through photolithography, dry etching, and the reflow process \cite{51}. First, the photoresist disks with an anticipated diameter are patterned on a silicon wafer with a 2-$\mu$m-thermal-oxide silica layer by photolithography. Second, the photoresist patterns are transformed to silica disks by buffered HF etching. Third, the silicon pillars underneath the silica disks are generated by isotropic XeF$_2$ dry etching. Finally, the microdisks are collapsed into toroids with ultrasmooth surfaces by CO$_2$ laser pulses, which melted the peripheries of the silica disks. The principle and minor diameter of the toroidal microcavity used in the experiment are 100 and 6 $\mu$m, respectively.

\emph{Gold nanoparticle solution.} The gold nanorod solution is chemically synthesized using a seed-mediated method following Ref. \cite{52}. The resulting nanorods are relatively uniform in sizes with an average length and diameter of 40 and 16 nm, respectively, exhibiting a longitudinal plasmon resonance around 680 nm confirmed by an extinction spectrum test [Fig. \ref{fig1}(a)]. The concentration of the particle solution is about 200 fM. The nanorods are naturally apart from each other due to surface charge \cite{52}.

\emph{Taper-toroid coupling system.} The coupling system of the taper and toroidal microcavity is illustrated in Fig. \ref{fig1}(b). A fiber taper of submicron waist diameter is fabricated by stretching a standard single mode fiber (Model SM600, Thorlabs). Two tunable lasers at a wavelength band of 680 nm (Model TLB6309, New Focus) and 635 nm (Model TLB6704-P, New Focus) are used to excite WGMs of the microtoroid via the evanescent coupling with the taper. The laser power is set as low as $\sim$100 $\mu$W to reduce the thermal effect. TE or TM modes are selectively excited using a polarization controller. A microfluidic channel is designed to immerse the cavity in gold nanoparticle solution, and part of the particle binding events can be observed from the scattering light using a CCD camera. The transmission light of the taper is collected by a low-noise photodetector (Model 1801, New Focus) and analyzed by an oscilloscope (Model: DLM2034, Yokogawa). The linewidth change $\Delta \kappa_T$ and the mode shift $\delta f$ can be obtained from the real-time fitting with an iteration of weighted least squared regression.

The coupling strength of the taper and the microcavity depends strongly on their relative position, and can be controlled by adjusting the taper position along both the radial ($\hat{r}$) and polar ($\hat{\theta}$)  directions of the cavity [shown in Fig. \ref{fig1}(c)]. At an optimal position, the decay rate equals to the coupling strength, and the transmitted light vanishes, corresponding to the critical coupling regime. When the taper is placed in the equator plane (i.e., $\theta = 0$), a gap from the cavity is necessary to achieve critical coupling, and the fiber taper suffers from oscillations induced by environment fluctuations, leading to experimental noises. Note that the WGM field confined in the toroid decays not only in the radial direction but also in the polar direction of the toroid [see finite element method (FEM) simulation in Fig. \ref{fig1}(c)]. While the taper is placed above or below the equator plane (i.e., $\theta \neq 0$), the critical coupling point can be achieved even when the taper attaches to the microtoroid surface, significantly reducing taper vibrations. In the experiment, we therefore position the taper at the $\theta\neq0$ location, working around the critical coupling point, to sustain both the high-$Q$ factor and the distinguishable WGM dip from the transmission signal.

\section{Results and Discussion}
The experimental uncertainties of the linewidth change and the shift of the cavity mode are firstly evaluated, when the microcavity system is immersed in deionized water. The intrinsic $Q$ factor of the WGM is $2\times10^7$ 
at the $680$-nm wavelength band [see the transmission spectrum in the inset of Fig. \ref{fig1}(d)]. The minor asymmetry in the line shape is resulted from the thermal effect of the laser which introduces extra experimental noise. The experimental uncertainty of the linewidth change and mode shift, using one standard deviation, are measured to be 1.35 and 1.96 MHz, respectively, in Figs. \ref{fig1}(d) and \ref{fig1}(e). The measured experimental uncertainty of the mode shift (1.96 MHz), using one standard deviation, is $\sim$1.5 times larger than that of the linewidth change (1.35 MHz), as a result of the laser jitter and an external drift in the resonance frequency [Fig \ref{fig1}(e)]. As for the mode shift, an extra step of background subtraction is applied to calculate the uncertainties, since evident frequency drifts of $-13.1$ MHz/s in deionized water are clearly seen in these data due to environmental temperature drift [see Fig. \ref{fig1}(e) inset]. Similar drift is also observed in another experiment \cite{24}. This linear frequency drift is removed from the raw data, and the generated data are applied to the analysis of the linewidth change and the mode shift.

\begin{figure}
\includegraphics[width = 8cm]{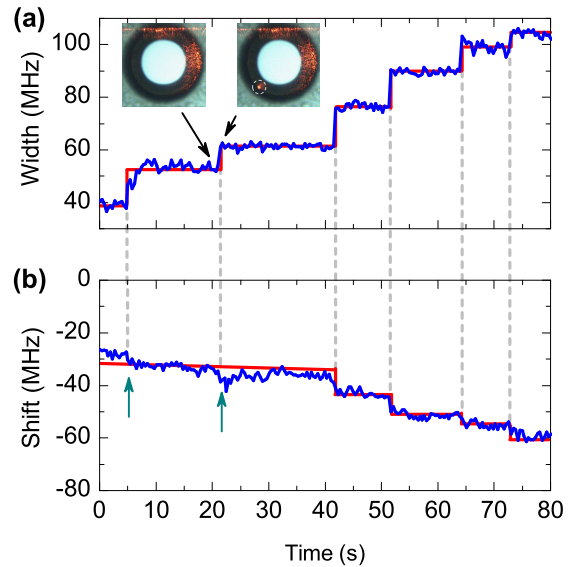}
\caption{Measured linewidth (a) and shift (b) of the WGM shown in Fig. \ref{fig1}(d) when the probe wavelength is ~$680$ nm (the plasmonic resonance of the particles). The microcavity is placed in the gold nanoparticle solution ($200$ fM). The gray dashed lines illustrate nanoparticle binding events at the same moment. The inset shows two top-view images, captured by a CCD camera, of the microcavity with a major (minor) diameter of $100$ $\mu$m ($6$ $\mu$m) before and after one binding event. The scattering light in the dashed circle indicates the newly bound particle.}
\label{fig2}
\end{figure}

To study the ability of the dissipative sensing method, the microcavity system is immersed in the solution of gold nanoparticles dissolved in deionized water with the concentration of 200 fM. The probe wavelength on the SPR around 680-nm band is used to excite WGMs, where the gold nanorods show the maximum extinction coefficient. Both the linewidth $\Delta \kappa_{T}$ and resonance shift $\delta f$ are plotted in Fig. \ref{fig2}, with the gray vertical dashed lines illustrating the single-particle-binding moments, and the red-guided lines are determined by a step-finder algorithm. Part of the particle binding events on the cavity are observed via the scattering light [see the top-view images of the microtoroid in the inset of Fig. \ref{fig2}(a) as an example, where the particle attached to the cavity is indicated in a dashed circle]. Note that single particles may have the chance to attach to the fiber taper, but those events only result in a power decrease in the transmission spectrum, and have no contributions to the linewidth change and mode shift. Evidently, as a response of the single-particle binding event, the linewidth and resonant wavelength could alter simultaneously in Fig. \ref{fig2}.
The threshold of the step changes is set as 3 times one standard deviation [i.e., experimental uncertainty $\sigma$ in Figs. \ref{fig1}(d) and \ref{fig1}(e)]. This $3\sigma$ degree of certainty is also used in other experimental data processing \cite{63,Noise}. On the other hand, a lower threshold could be set such that two minor step changes emerge at t $\sim$ 5 and 21 s in Fig. 2(b), in which case, however, some other unreliable changes such as spike noises may also be misread as single nanoparticle binding events.
As expected, the response is distinct from each other at some points. For example, a few single-particle binding events, arrows pointed in Fig. \ref{fig2}(b), cannot be distinguished by the mode shift measurement, because the shifts are below the threshold of the step changes. This phenomenon is caused by the fact that the real part of the polarizability of a metal particle approaches zero when the frequency is on SPR. In this case no evident shift of the WGM can be observed due to the absence of backscattering, while the linewidth change can still respond to the particle binding events due to absorption. Additionally, the nonzero imaginary part of the polarizability contributes to the side scattering instead of the backscattering, permitting one to observe the particle through the CCD camera in the inset of Fig. \ref{fig2}(a).

\begin{figure}
\includegraphics[width = 8cm]{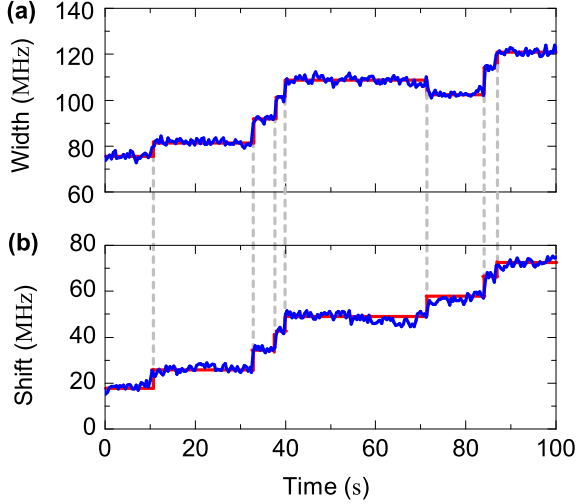}
\caption{Measured linewidth (a) and shift (b) of the WGM when the probe wavelength is ~$635$ nm. The probe wavelength is off the surface plasmon resonance of the particles. Other experimental conditions are the same with that in Fig. \ref{fig2}.}\label{fig3}
\end{figure}

The laser probe light is then swept around 635 nm which is off-SPR wavelength to weaken the nanoparticle absorption, and similarly, the linewidth and mode shift over 100 s are shown in Fig. \ref{fig3}. Different from Fig. \ref{fig2}, all observed nanoparticle binding events can be identified simultaneously via both the reactive and dissipative sensing methods. As expected, the mode shift magnitude at 635 nm increases compared to that in the $680$-nm case, because the real part of polarizability of a single particle becomes nonzero, and the backscattering of WGMs adds to the mode shift. A linewidth decrease appears at $70$ s, resulting from the multiparticle interference of the nanorods studied in Refs. \cite{46,47,48}.

\begin{figure}
\includegraphics[width = 8cm]{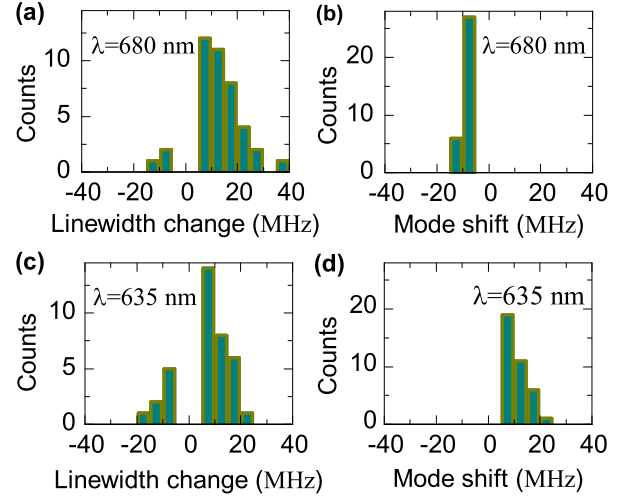}
\caption{Distributions of linewidth change and mode shift induced by gold nanorods binding on the microtoroid in 10 min. The probe wavelength is $\sim$680 nm [(a) and (b)] or $\sim$635 nm [(c) and (d)].}\label{fig4}
\end{figure}

The distributions of the linewidth change and mode shift above 3 times the uncertainties are plotted in Fig. \ref{fig4}. From the statistical analysis, the dissipative sensing measurement catches eight more nanorods binding events than the reactive sensing method at 680-nm probe wavelength in 10 min [Figs. \ref{fig4}(a) and \ref{fig4}(b)]. Furthermore, the average linewidth change ($\sim$13.5 MHz) at 680 nm is $\sim$1.3 times larger compared to the value ($\sim$10.2 MHz) at 635 nm due to the SPR-enhanced field inside the nanoparticle. Inversely, the average mode shift ($\sim$11.2 MHz) at 635 nm is exhibited at $\sim$1.7 times larger than the value ($\sim$6.5 MHz) at the on-resonance wavelength due to the nonzero real part of the polarizability as discussed above. The signal-to-noise ratio (SNR) of the linewidth change and mode shift are $\sim$3.3 and $\sim$1.1 at 680 nm, respectively, while the values become $\sim$2.5 and $\sim$1.9 at 635 nm. In particular, the SNR of the mode shift on SPR is as low as $\sim$1.1, which explains the difficulty in observing single-particle binding events via the reactive sensing method. Remarkably, the SNR of the linewidth change as a result of dissipative interactions remains higher than that of the mode shift at both probe wavelengths; for example, it is about 3 times higher on SPR, thus providing a better sensing performance, compared with the conventional reactive sensing method when the mode shift is almost unresolvable from the background noise.

The experimental results above can be verified theoretically according to the Wigner-Weisskopf semi-quantum-electrodynamics treatment. In theory, the subwavelength particle positioned on the microtoroid surface is polarized by the evanescent field of the WGM which can be considered to be uniform. The total loss induced by the particle originates from the side scattering and absorption loss, i.e.,
\begin{equation}
   \gamma = \gamma_s+\gamma_a,
\end{equation}
in which
\begin{equation}
  \gamma_s = \frac{\varepsilon_m^{5/2}|\alpha|^2 f^2(\vec{r})\omega_c^4}{6\pi c^3 \varepsilon_c V_c}
\end{equation}
   and
\begin{equation}
    \gamma_a =\frac{ \text{Im}[\varepsilon]f^2(\vec{r})\omega_c V_p|\beta|^2}{\varepsilon_c V_c}.
 \end{equation}
 Therefore, the linewidth change is determined by both the real and imaginary part of the polarizability, which is similar as the interferometric scattering (iSCAT) microscopy \cite {61}. Here, $\varepsilon$ and $\alpha$ are the permittivity \cite{49} and polarizability of the metal particle, $\varepsilon_m$ and $\varepsilon_c$ are the permittivity of the surrounding medium and the cavity, $f(\vec{r})$ denotes the cavity mode function, $\omega_c$ represents the degenerate angular frequency of CW and CCW WGMs, $V_c$ and $V_p$ are the cavity mode volume and the particle volume, $\beta$ is the root-mean-square enhancement of the electric field inside the nanorod which can be calculated from FEM simulation, and $c$ is the speed of light in vacuum. In addition, the backscattering coupling strength, contributing to the mode shift, can be written as
 \begin{equation}
  g = -\frac{\varepsilon_m \text{Re}[\alpha]f^2(\vec{r})\omega_c}{2\varepsilon_c V_c},
 \end{equation}
 which is only dependent on the real part of the polarizability.

\begin{figure}
\includegraphics[width = 8cm]{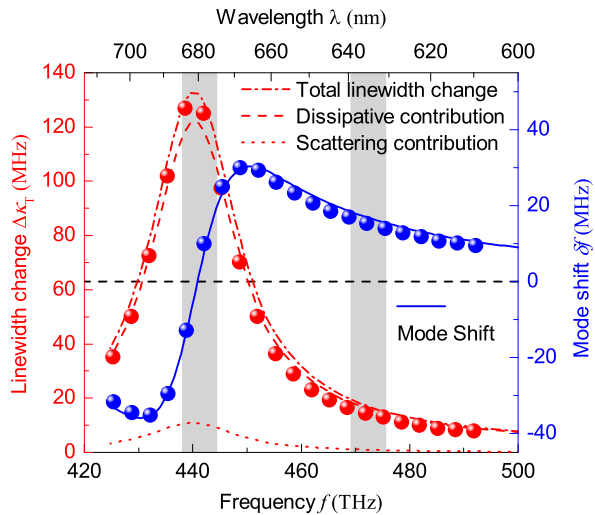}
\caption{Theoretical predictions of linewidth change $\Delta\kappa_T$ (left axis, dash-dotted curve) and mode shift  $\delta f$ (right axis, solid curve) at different probe light frequency (wavelength) for a gold nanorod (40 nm $\times$ 16 nm) bound in a WGM with $Q=2\times10^7$ , $V_c\approx450\ \mu\text{m}^3$ and $f^2(\vec{r})\approx0.2$. The dissipative interaction (dashed curve) contributes 10 times higher to the linewidth change than the nanoparticle scattering (dotted curve) does. The symbols represents the linewidth change and mode shift using the 3D model via {\fontsize{7pt}{12pt}\selectfont COMSOL} software.}\label{fig5}
\end{figure}

As shown in Fig. \ref{fig5}, the linewidth change induced by the dissipative interaction (dashed curve) is more than 10 times higher than that by the scattering interaction (dotted curve). Therefore, $\Delta \kappa_T$ (dash-dotted curve) mainly results from the particle absorption and differs from the backscattering-induced mode broadening. When the probe wavelength is about 680 nm (SPR wavelength), the linewidth change induced by a single metal nanoparticle reaches a maximum, while the mode shift remains close to zero. This theoretical calculation explains the experimental results in Fig. \ref{fig2}, where some binding events can be distinguished only via the linewidth change and the reactive sensing would fail. In Fig. \ref{fig5}, the mode shift induced by the reactive interaction vastly increases when the wavelength is detuned from SPR. For example, around 635 nm, $\delta f$ reaches $16.8$ MHz, which is far from zero. As the wavelength increases, the real part of the particle polarizability turns from negative to positive values. When  $\text{Re}[\alpha]=0$ at a wavelength close to SPR, the dipole moment of the nanoparticle has a $\pi/2$ phase difference with the evanescent field, and the coupling energy between the nanoparticle and the cavity is zero, resulting in a zero mode shift \cite{si}. The positive mode frequency shift in Fig. \ref{fig5} appears corresponding to the prediction when $\text{Re}[\alpha]<0$. Although in this case, gold nanorods experience a repulsive optical force from the optical field \cite{50}, the positive frequency shifts are still observed in Figs. \ref{fig3}(b) and \ref{fig4}(d), possibly due to the electric force between the negatively charged silica cavity and the positively charged gold nanorods. Note that the WGM wavelength in Fig. \ref{fig2} is $682.8$ nm, thus, no positive mode shift is observed, which is consistent with the theoretical predictions in Fig. \ref{fig5}. The 3D finite-element-method simulations using {\fontsize{8pt}{12pt}\selectfont COMSOL} 4.3a rf module is also conducted to verify the theoretical predictions in Fig. \ref{fig5}, and the detailed simulation conditions can be found in Ref. \cite{si}. Although the magnitudes of the linewidth change and the mode shift do not fall exactly on the experimental results due to the nonuniform size distribution of the the nanorods as well as their random position on the cavity and orientation, the general trend is the same, for example, the linewidth change is larger than the mode shift at 680 nm, and vice versa at 635 nm \cite{si}.

Theoretically, the largest linewidth change induced by a $40$ nm $\times16$ nm gold nanorod is 130 MHz at the SPR wavelength, i.e., $680$ nm (Fig. 5). Considering the threshold of finding a linewidth step change is $4.05$ MHz (i.e., 3 times the experimental uncertainty), the smallest dimension of a nanorod that can be detected by the dissipative sensing method is about $13$ nm $\times$ 5 nm at its SPR wavelength. Similarly, the minimum detectable dimension of a gold nanorod is $23$ nm $\times$ 10 nm at the wavelength when the mode shift peaks maximum. Thus, the linewidth change method can detect a single gold nanorod of $\sim$$12$ times smaller volume than that observed by the mode shift, representing a better detection limit.

While the microscopy techniques, such as dark field, iSCAT \cite{58,61}, and photothermal microscopy \cite{54}, which achieve the imaging of single analytes in a high speed, microcavity sensing has less restriction on the stability of the light source, and is ready for on-chip integration. In addition, the tunable laser or a high-resolution spectrometer in the conventional microcavity sensing setups could be replaced by other types of coherent measurement methods such as the microcavity lasing beat note \cite{34,36,37} or cavity ring-up spectroscopy \cite{12}.

\section{Conclusion}

In summary, a dissipative sensing method is demonstrated to detect single lossy nanoparticles with a high-$Q$ toroidal microcavity. By monitoring both the linewidth change and mode shift simultaneously in the experiment, the performance of the dissipative sensing shows a better detection ability compared to the general reactive sensing method, and the detection of single gold nanorods is realized in the aqueous environment. Tuning the probe wavelength on SPR, the mode shift of WGMs approaches zero due to the absence of backscattering caused by the zero real part of the particle polarizability, while the dissipative interaction still works. The experimental results fall within the theoretical prediction. The dependence of the linewidth change and the mode shift on frequency may provide a way to obtain the information of the nanoparticles in the electric field, such as the polarizability of the particle. This dissipative sensing method holds great potential in detecting nanoparticles of high absorption or ultralow polarizabilities, such as carbon nanotubes and metal nanoparticles, and in characterizing nanoparticle properties in combination with the reactive sensing method.

\section{Acknowledgment}

Y.F.X. thanks Stephen Arnold, Lan Yang, and Xue-Feng Jiang for stimulating discussions. Y.F.X. provided the idea, designed the experiment, and supervised the project. This work is supported by the 973 Program (Grants No. 2013CB921904 and No. 2013CB328704), the NSFC (Grants No. 61435001, No. 11474011, and No. 11222440), and the Beijing Natural Science Foundation Program (Grant No. 4132058). B. Q. S. was supported by the National Fund for Fostering Talents of Basic Science (Grants No. J1030310 and No. J1103205).

B. Q. S., X. C. Y. and Y. Z. performed the experiment, analyzed the data, and contributed equally to this work. All
authors contributed to the discussion and paper writing.


\end{document}